\documentstyle[psfig]{article}
\begin{document}

\centerline {\bf On the Origin of Nuclear Superfluidity}

\vskip 1cm 

{F. Barranco $^{a)}$, 
R.A. Broglia $^{b),c),d)}$,
G. Gori $^{b)}$, E. Vigezzi$^{c)}$,
P.F. Bortignon$^{b),c)}$,  J. Terasaki $^{c)}$

\vskip 5mm

\parindent =0pt 

a) Escuela de Ingenieros Industriales, Universidad
de Sevilla, Camino de los Descrubimientos, Sevilla, Spain

b) Dipartimento di Fisica, Universit\`a di Milano,
Via Celoria 16, 20133 Milano, Italy

c) INFN Sez. Milano, Via Celoria 16, Milano, 20133 Italy

d) The Niels Bohr Institute, University of Copenhagen,  
Blegdamsvej 17, DK-2100 Copenhagen, Denmark

\parindent= 20pt 

\vskip 2cm

\centerline {\bf ABSTRACT}

\vskip 5mm

         The induced pairing interaction arising from the exchange of 
collective surface vibrations among nucleons moving in time reversal 
states close to the Fermi energy is found to lead to values of the
pairing gap which are similar to those experimentally observed.

\vskip 2cm

      It is well established that nucleons moving close to the Fermi energy 
in time reversal states have the tendency to form Cooper pairs which 
eventually condense\cite{bohr1},\cite{bohr2}. This phenomenon, which parallels that which is 
at the basis of low temperature superconductivity \cite{[3]}, modifies in an important way the 
nuclear properties, in particular the occupation number of single-particle
levels around the Fermi energy, the moment of inertia of deformed nuclei, the
lifetime of cluster decay and of fission process, the depopulation
of superdeformed configurations, the cross section of
two-nucleon transfer reactions, etc. 
(cf. e.g. refs. \cite{bohr2},\cite{[4]}-\cite{[16]} and refs. therein).

     While in the case of superconductivity the attraction among the electrons
is generated by the exchange of phonons, in the nuclear case the origin of 
the pairing interaction is related to the $^1S_0$ phase shift of 
the free nucleons, which is attractive at low relative momenta.
 In keeping with the fact that the free nucleon-nucleon interaction is
strongly renormalized in nuclei, Cooper pair formation can benefit 
from the exchange of low-lying collective surface vibrations between pairs of 
nucleons moving in time reversal states close 
to the Fermi energy (cf. ref. \cite{bohr2}, p.432 and refs. \cite{[17]},
\cite{[18]}). 
In fact, we shall show below that 
this mechanism gives rise to pairing gaps which are similar to those
observed experimentally.

Calculations have been carried out for a number of isotopic chains:
$^{A}_{20}$Ca, $^A_{22}$Ti  and $^{A}_{50}$Sn. The results provide insight
into the role the induced interaction plays in neutron and proton pairing
correlations in nuclei. 
Calculations have also been carried out for the case of $^{42}_{21}$Sc 
and found to lead to strong proton-neutron pairing correlations.

The properties of the collective surface vibrations were determined  within 
the framework of the random-phase approximation and of the particle-vibration
coupling model (cf. e.g. refs. \cite{bohr2},\cite{[19]}). 
The induced interaction (cf. Fig. 1 (inset))
$$
 < \nu' \bar{\nu'} |v| \nu \bar{\nu} > = 
\sum_{\lambda n} 
\frac {<\nu||R_o \frac {\partial U}{\partial r} Y_\lambda ||{\nu'}>|^2}
{\sqrt{(2j_\nu+1)(2j_{\nu'}+1)}} 
\frac{2 (2 \lambda+1) \Lambda_{\lambda}^{2}(n)}
{E_o - (\epsilon_\nu + \epsilon_{\nu'} + \hbar \omega_{\lambda} (n))},
\eqno(1)
$$
was then calculated, and the state dependent pairing gap
$$
\Delta_{\nu} = - \sum_{\nu'}  \frac{\sqrt{2j_{\nu'} +1}}{\sqrt{2j_{\nu}+1}}
\frac {\Delta_{\nu'}}{2E_{\nu'}}
<\nu' \bar{\nu'} |v| \nu \bar{\nu} >, \eqno(2)
$$
determined.
In Eq.(1), 
the quantum  numbers of the single-particle states are labeled by
$\nu (\equiv n_{\nu},l_{\nu},j_{\nu})$.
The states $| \nu \bar \nu>$ are coupled to angular momentum zero, and
$\bar \nu$ denotes the state time reversed to $\nu$. The corresponding
single-particle energies are denoted $\epsilon_{\nu}$ and $\epsilon_{\nu'}$. 
The quantity $E_{o}$ is the energy of the resulting ground state,
in keeping with Bloch-Horowitz perturbation theory \cite{[20]}. 
The quantity inside the reduced matrix element is the 
particle-vibration formfactor (deformation potential \cite{[19]}, see also
ref. \cite{[21]}), product of the
nuclear radius $R_o$, of the derivative $\partial U/\partial r$ of the
single-particle potential and of a spherical harmonic of multipolarity 
$\lambda$. The potential $U$ is assumed to have Woods-Saxon shape, and is
parametrized according to ref. \cite{[22]}. The quantities $E_{\nu'}$ in Eq.(2)
denote the quasiparticle energies.

The
quantity $\Lambda_{\lambda} (n) = \frac{\beta_{\lambda}(n)}
{2 \lambda +1} $ 
is the particle-vibration coupling strength
associated with the n-th vibration (in order of increasing energy)
of multipolarity $\lambda$ and frequency
$ \omega_{\lambda}(n)$. 
The values of the coupling strength for the even- and odd-multipolarities 
have been determined  so as to reproduce the experimental  transition 
probabilities of the low-lying $2^+ $ and 3$^-$ vibrational 
states.
The calculations have then been carried out making use
of vibrational states of multipolarities and parities 
$\lambda^{\pi} =2^+,3^-,4^+$ and $5^-$.

In Fig. 1 we show the calculated state dependent pairing gap 
for the nucleus $^{120}$Sn. For levels close to the Fermi energy, this quantity
is of the order of 1 MeV. This result is to be compared with the empirical
value of 1.5 MeV, obtained making use of the relation

$$
\Delta= \frac{1}{2} \left [ B(N-2,Z) + B(N,Z) - 2 B(N-1,Z) \right ],
\eqno(3)
$$
where $B(N,Z)$ is the binding energy of the nucleus with $N$ neutrons and
$Z$ protons.

In Fig. 2, we show the value of the state dependent pairing gap averaged 
over levels lying within an energy interval of the order of 
$ \pm 2 \Delta$ around 
the Fermi energy, for a number of  Sn-isotopes in comparison with 
the corresponding values obtained from  Eq. (3). 
In all cases, theory accounts for a  consistent fraction of the empirical 
values of the pairing gap.

In Fig. 3 we display the results of calculations carried out for the isotopes
$^{A}$Ca and $^{A}$Ti, in comparison with the corresponding 
results of Eq.(3). As in
the previous case, the induced interaction leads to pairing gaps
which account for a large fraction of the empirical value, 
and which furthermore display a similar dependence with $A$, a behaviour
which reflects the fluctuations of the nuclear surface. In particular,
the low predicted value of $\Delta$ in $^{50}$Ca as compared to $^{42}$Ca
is due to the fact that the "core" $^{48}$Ca is more rigid than the "core"
$^{40}$Ca. 
We have also determined the induced proton-neutron pairing
interaction in $^{42}$Sc, arising  from the exchange of the low-lying
collective surface vibrations of the core $^{40}$Ca. The calculated value
of 1.5 MeV (cf. also the result obtained for $^{42}$Ca, Fig. 2)
essentially coincides with the empirical value (1.6 MeV) obtained
making use of Eq. (3).

We conclude that the exchange of low-lying surface 
vibrations among nucleons moving in time reversal states close to the Fermi
energy, gives rise to an induced pairing interaction which leads to pairing
gaps of similar magnitude to those experimentally observed. This result will force
us to review our present understanding of the pairing phenomenon in
nuclei, in particular its description in terms of $\omega-$independent
effective interactions. It will also have consequences in the
analysis of phenomena like the quenching of the pairing gap
taking place as a function of the angular momentum and of the energy 
(temperature) content of the nuclear system.
In keeping with these results, and because collective vibrations couple
democratically to all nucleons, regardless of the isospin quantum number,
the induced interaction mechanism is expected to lead to a conspicuous
proton-neutron pairing correlation.

\vskip 2cm 

{\bf Captions to the figures}

\vskip 1cm

\underbar{Fig. 1}

State-dependent pairing gap $\Delta_{\nu}$ (cf. Eq.(2)) for the nucleus
$^{120}$Sn, calculated making use of the induced interaction defined in
Eq.(1) (cf. inset, where particles are denoted by arrowed lines and phonons
by a wavy line).

\vskip 1cm

\underbar{Fig. 2}

Average value of the state dependent pairing gap associated with levels
lying close to the Fermi energy of $^A_{50}$Sn-isotopes, calculated as
discussed in the text, making use
of the pairing gap defined in Eq.(2),
in comparison with the empirical pairing gap (cf. Eq.(3)). The results of 
two calculations are shown, 
associated with RPA solutions which fit two different sets of transition 
probabilities  connecting the lowest lying quadrupole and octupole vibrations
with the ground state. The first set (also used in the calculation shown
Fig. 1 for $^{120}$Sn) 
was taken from ref. [23], while the second
set from ref. [24]. The corresponding theoretical results are denoted by solid
squares and triangles respectively.}

\vskip 1cm

\underbar{Fig. 3}

Average value of the neutron pairing gap 
of the $^{A}_{20}$Ca-isotopes and of the proton pairing gap of the 
$^{A}_{22}$Ti-isotopes, in comparison
with the empirical pairing gap (cf. Eq.(3)). In both cases the pairing matrix
elements have been obtained from Eq. (1),
making use of the vibrational states of the "core" $^{A-2}$Ca, and using
the values of the experimental transition probabilities given in ref. [25].
The gap of the Ca-isotopes has been calculated making use of Eq.(2).
In the case of the
Ti isotopes, the matrix of the induced interaction  was diagonalized.
The pairing gap was then calculated making use of the relation 
$\Delta= \frac{1}{2} \left [ B(N,Z-2) + B(N,Z) - 2 B(N,Z-1) \right ]$,
the proton analogous to the expression given in   Eq.(3)  and used to calculate 
the neutron pairing gap.

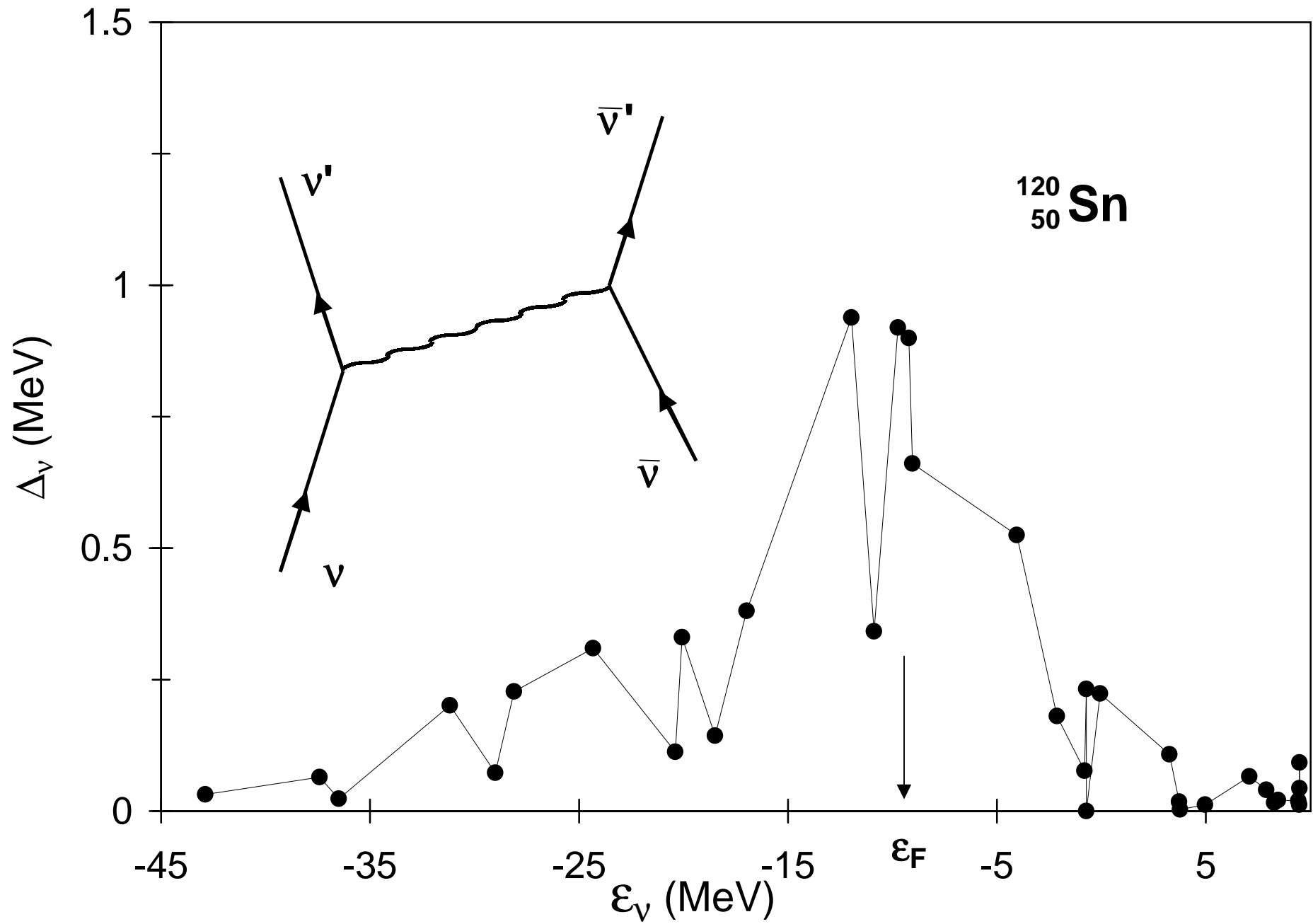
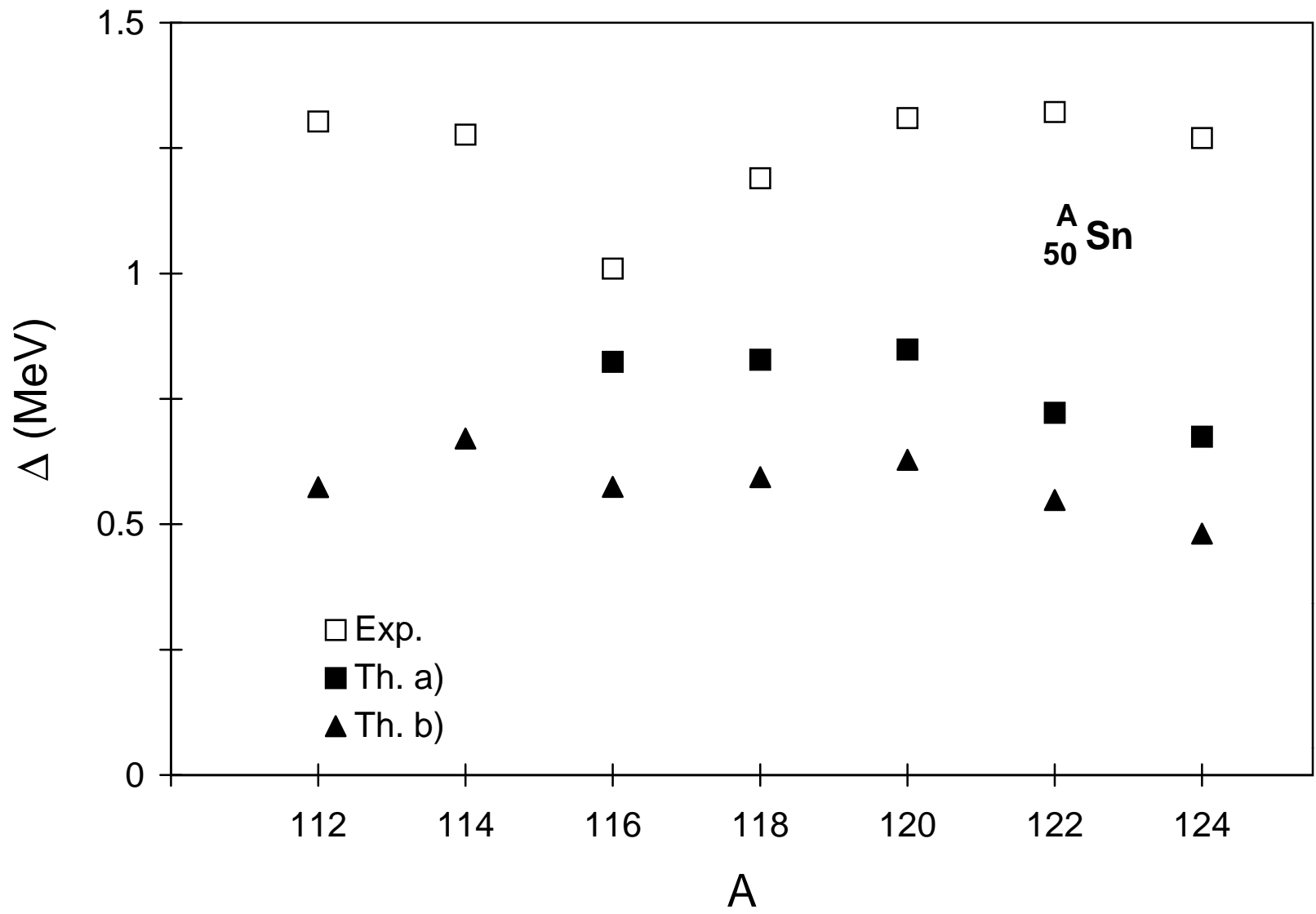
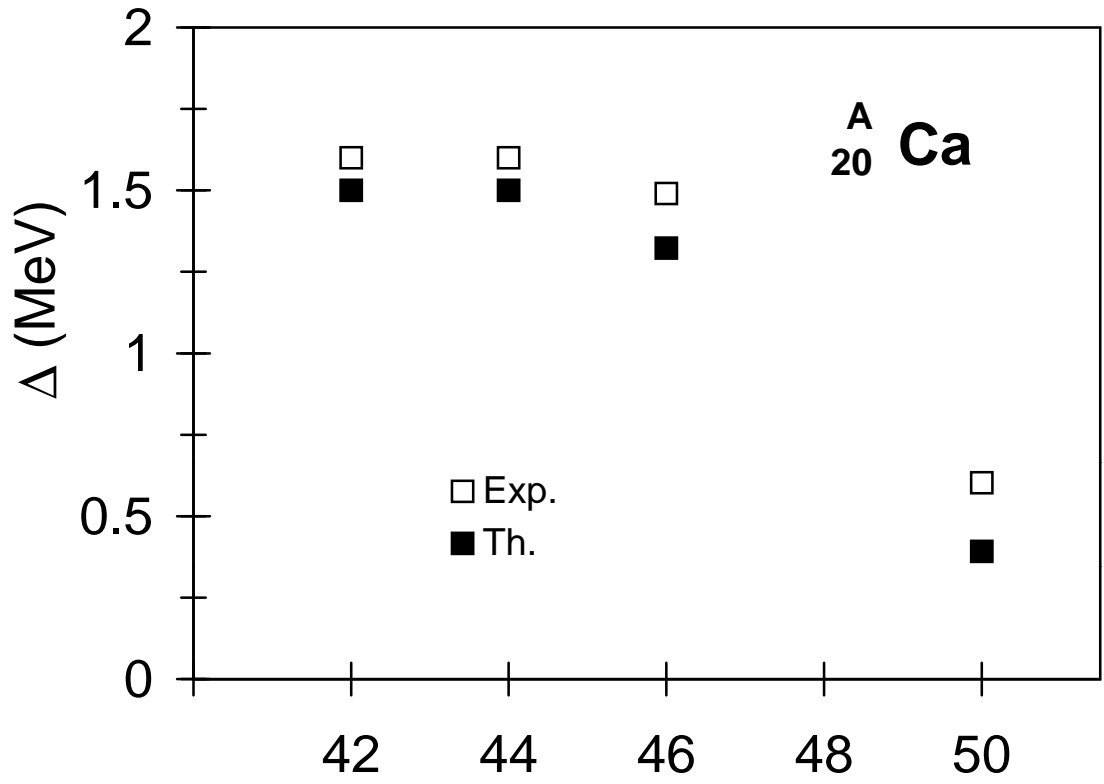
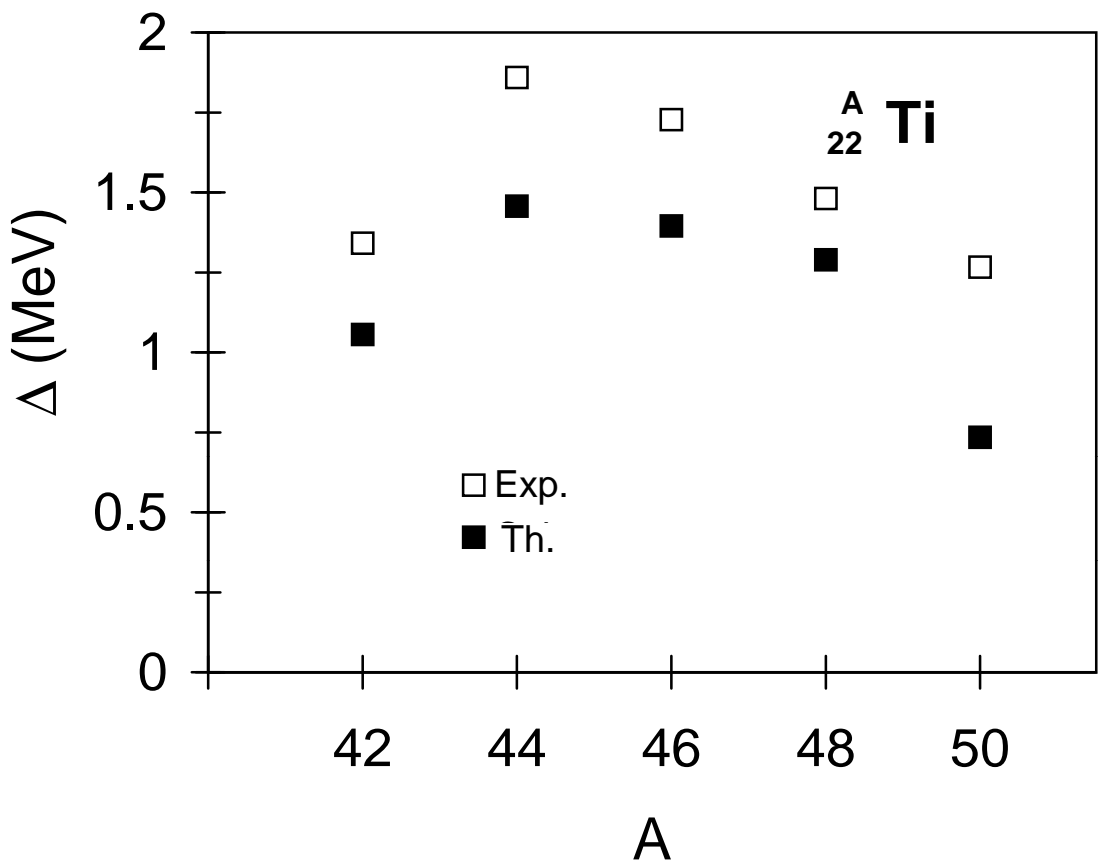
\end{document}